\documentclass{collective-intelligence} 
\usepackage{synttree}
\usepackage[hang]{subfigure}

\newcommand{\avg}[1]{\left< #1 \right>} 
\renewcommand{\d}[2]{\frac{d #1}{d #2}} 

\begin{document}
\bibliographystyle{agsm}

\title{EFFECTS OF SOCIAL INFLUENCE ON THE WISDOM OF CROWDS }
%
%
%
%
%

\numberofauthors{3} 
%
\author{
%
%
  \alignauthor
  Pavlin Mavrodiev \\
  \affaddr{Chair of Systems Design}\\
  \affaddr{ETH Zurich}\\
  \affaddr{Kreuzplatz 5, CH-8032 Zurich}\\
  \email{pmavrodiev@ethz.ch}
  \alignauthor
  Claudio J. Tessone\\
  \affaddr{Chair of Systems Design}\\
  \affaddr{ETH Zurich}\\
  \affaddr{Kreuzplatz 5, CH-8032 Zurich}\\
  \email{tessonec@ethz.ch}
  \alignauthor
  Frank Schweitzer\\
  \affaddr{Chair of Systems Design}\\
  \affaddr{ETH Zurich}\\
  \affaddr{Kreuzplatz 5, CH-8032 Zurich}\\
  \email{fschweitzer@ethz.ch}
  \alignauthor
}

\maketitle
\begin{abstract}
  Wisdom of crowds refers to the phenomenon that the aggregate prediction
  or forecast of a group of individuals can be surprisingly more accurate
  than most individuals in the group, and sometimes -- than any of
  the individuals comprising it. This article models the impact of social
  influence on the wisdom of crowds. We build a minimalistic
  representation of individuals as Brownian particles coupled by means of
  social influence.  We demonstrate that the model can reproduce results
  of a previous empirical study.  This allows us
  to draw more fundamental conclusions about the role of social
  influence: In particular, we show that the question of whether social
  influence has a positive or negative net effect on the wisdom of crowds
  is ill-defined. Instead, it is the starting configuration of the
  population, in terms of its diversity and accuracy, that directly
  determines how beneficial social influence actually is. The article
  further examines the scenarios under which social influence promotes or
  impairs the wisdom of crowds.
\end{abstract}

\section{Introduction}
\label{introduction}
Contrary to popular belief, the wisdom of crowds\footnote{The terms
  \textit{crowd} and \textit{group} are used interchangeably, as are
  \textit{opinions} $\leftrightarrow$ \textit{estimates}
  $\leftrightarrow$ \textit{judgements} $\leftrightarrow$
  \textit{beliefs} and \textit{configuration} $\leftrightarrow$
  \textit{state}.} is a statistical and
not a psychological phenomenon. Wisdom, in this context, refers to the
aggregate opinion of a population being closer to a true value than most
individual opinions. The idea of aggregating over a space of opinions
can, in an ergodic fashion, also be applied to aggregating over an
individual's own perspectives over time, with the same benefits -- i.e.~it 
may yield a more accurate decision making  \cite{Rauhut2011191}.

However, the wisdom of crowds is not a pure statistical regularity, in
the sense that more does not imply better. It necessitates certain
conditions, which can be summarised, following \cite{Surowiecki2005}, as
diversity and independence of opinions, specialisation in expert
knowledge and a mechanism for aggregating individual opinions. From this
perspective, the best collective decisions do not rely on consensus
building and compromises, but instead on aggregating many heterogeneous
views -- given enough diversity in opinions, the errors in each of them
cancel out until only useful information is left \cite[p.10]{Surowiecki2005}.

Diversity has been identified as instrumental in providing creative
perspectives to problem-solving, thus avoiding getting stuck in locally
suboptimal solutions \cite{Hong2004}. In fact, the ``diversity prediction
theorem''\footnote{The theorem states that collective accuracy equals
  average individual error minus the variance in opinions, i.e. group
  diversity.} \cite{Page2007} shows that diversity weighs as much as
individual ability in determining collective accuracy. From a
mathematical point of view, diversity is required in order to balance out
uncorrelated imperfections in opinions through aggregation. Intuitively,
as no single individual is aware of all traits of a given problem at
hand, diversity helps people combine their idiosyncratic perceptions so
that together they gain a wider perspective. Diversity, however, is not
the pinnacle of optimal decision-making. No amount of diversity can help
if the population is completely ignorant on a given issue, or if opinions
are diverse, but heavily skewed. Thus, the composition of diversity is as
important as diversity itself \cite{Bonabeau2009}.

Independence of opinions is another important distinguishing feature of
the wisdom of crowds for it either excludes communication, information
spreading, learning and social influence processes, such as herding and
imitation or limits the effect of such processes should they be
present. In this context, it is useful to draw a conceptual difference
between wisdom of crowds, as commonly understood, and ``collective
intelligence''. Wisdom of crowds is a quantification of the state currently
occupied by a given group, such as an aggregate opinion. Intelligence, pertains to the ability of individuals to
learn, to understand, and to adapt to arbitrary external conditions using
own knowledge \cite{Leimeister2010}. Collective, describes a group of
individuals pooling their intelligence together for a common
purpose. As such, collective intelligence can be seen
as the mechanism by which groups converge to a certain collective decision, whereas
wisdom of crowds is the numerical representation of the said decision. In
this paper, we are interested in how the collective intelligence
mechanism affects the wisdom of crowds.

Recent empirical evidence has shown that enabling collective
intelligence by introducing social influence, can be detrimental to the
aggregate performance\footnote{Performance is the pair
  \{$\mathcal{E}(t)$,$\mathcal{W}(t)$\}. See next section.} of a population \cite{Lorenz2011}. By
social influence, we understand the pervasive tendency of individuals to
conform to the behaviour and expectations of others \cite{Kahan2007}. In
separate experiments, \citename{Lorenz2011} asked participants 
to re-evaluate their opinions on quantitative subjects over several
rounds and under three information spreading scenarios -- no information
about others' estimations (control group), the average of all opinions
in each round and full information on other subjects'
judgements. They found evidence that under the latter two
regimes, the diversity in the population decreased, while the collective
deviation from the truth increased. This result justified the
disheartening conclusion that allowing people to learn about others'
behaviours and adapt their own as a response does not always lead to the
group acting ``wiser''\footnote{Wiser, in the numerical sense of the
  wisdom of crowds.}. Rather, as the authors posited, not only is the
population jointly convinced of a wrong result, but even the simple
aggregation technique of the wisdom of crowds is deteriorated. From a
policy-maker's perspective, such groups are, thus, not wise.

Current research has not yet investigated thoroughly the theoretical link
between social influence and its effect on the wisdom of crowds. In this
paper, we build upon the empirical study in \cite{Lorenz2011} by
developing a formal model of social influence. Our goal is to unveil
whether the effects of social influence are unconditionally positive or
negative, or whether its ultimate role is
mediated through some mechanism, so that the effect on the group wisdom
is only indirect. We adopt a minimalistic agent-based model, which
successfully reproduces the findings of the said study  and
gives enough insight to draw more general conclusions. In particular, we confirm that small amounts of
social influence lead to faster convergence, however, it is the starting
configuration\footnote{Configuration is the pair
  \{$\mathcal{E}(t),\mathcal{D}(t)$\}. See next section.} of the population (in terms of its initial diversity and
deviation from the truth) that ultimately attribute the net effect of
social influence on the wisdom of crowds.

The rest of the paper is organised as follows. The next section reviews the empirical study on which our model is
based, together with its main results and the measures used to quantify the collective performance of the groups. The model itself is presented
after that. Results and conclusions follow as the last two sections respectively.
\section{The Experiment}
\label{the-experiment}
One of the latest study of the effect of social influence on
wisdom of crowds aimed at quantifying how people's opinions are
influenced by others and to what extent this influence relates to the
aggregate deviation of the group from an objective truth \cite{Lorenz2011}.

The authors recruited 144 subjects among students at ETH Z\"{u}rich. The
subjects were split into 12 experimental sessions, each consisting of 12
participants. During each session the subjects were asked a total of 6
quantitative questions regarding geographical facts and crime
statistics\footnote{Example: What is the population density of
  Switzerland?}. Each question had to be answered over five time
periods. The questions were designed in such a way that individuals were not likely
to know the exact answer, but could still formulate educated guesses.

\begin{figure}[htbp]
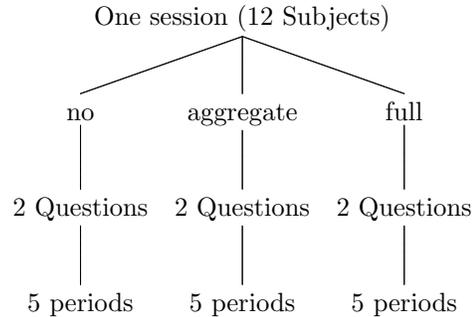

  \centering \synttree[ One session (12 Subjects) [no [2 Questions [5
  periods]] ][aggregate [2 Questions [5 periods]]][full [2 Questions [5
  periods]]] ]
  \caption{Experiment structure}
  \label{fig-experiment-structure}
\end{figure}

Figure \ref{fig-experiment-structure} shows the structure of the
experiment.

In the first time period, all subjects responded to the particular
question on their own. After all 12 subjects gave their estimations, each
had to answer the same question over 4 additional periods for a total
of 5 rounds. Three different information conditions were tested
regarding the information that participants learnt about the answers of
others in the previous time periods. In the ``no'' information regime, no
individual was aware of others' opinions throughout all 5
periods. In the ``aggregate'' regime, each subjects was provided with the
arithmetic average of everyone else's answers in the previous
round. Finally, in the ``full'' information regime, subjects learnt all
opinions from all previous rounds so far. In each session, two questions
were posed in the no, two in the aggregated, and two in the full
information condition. The 12 subjects were randomly assigned to the
three information conditions in the beginning of the session.

An important component of the experiment was the reward
structure. Participants received individual awards depending on their
deviations from the correct answer -- reward-bearing deviations were
defined as 10\%, 20\% and 40\% intervals around the truth. The rewards
were provided at each round, so that individuals were motivated to make
 optimal decisions at all times. However, the correct answer and rewards were disclosed at the
end of the experiment to avoid giving away a-priori knowledge about the truth. This
reward structure eliminated the benefits of strategic considerations,
such as misleading or cooperating with others, because it did not affect
individuals' payments -- subjects had incentives to
use only their own knowledge and interpretation of others' opinions to
find the truth.

As we mentioned above, a requirement for quantifying the wisdom of crowds
is a suitable aggregation measure. Ideally, an aggregated measure would
indicate the most central opinion in the population. For opinion
distributions that are Gaussian-like, the simple unweighted arithmetic
average may be a good choice. There are other averaging methods,
which could perform better in special cases
\cite{Genest1986,Dawid1995,Hegselmann2005}. The opinion distributions in the experiment
were found to be heavily right-skewed with a majority of low estimates
and a minority spread on a fat right tail, much like log-normal
distributions. The skew made the arithmetic average inappropriate as a
measure of centrality, since it was closer to the truth than individuals'
first estimates in only 21.3\% of the cases. However, taking the natural
logarithm of the estimates, resulted in a more bell-shaped
distribution. Consequently, the arithmetic average of the log-transformed
opinions (which equals the logarithm of the geometric mean of the
original data) performed much better -- it was closer to the true value
than individuals' first estimates in 77.1\% of the cases. This provided a
justification for log-transforming all opinions over all sessions before
further analysis\footnote{The authors in fact took the logarithm
  of the original estimates divided by the true values, so that answers
  across different questions could be compared}.

Three main quantities were used to evaluate the aggregate performance of
the crowd -- ``collective error'', ``group diversity'' and ``wisdom of
crowds indicator''. Let $N$ be the number of individuals and $x_{i}(t)$
be the answer of individual $i$ in round $t$. The collective error,
$\mathcal{E}(t)$, is defined as the squared deviation of the average opinion
in round $t$ from the true value, $\mathcal{T}$. Since all $x_{i}$'s were
log-transformed, the arithmetic average is $\avg{\ln x(t)}$\footnote{The
  notation $\avg{X}$ stands for $(1/N)\sum_{i=1}^{N}X_{i} $}, which
equals the logarithm of the geometric mean of the original data, thus the
collective error in period $t$ equals:
\begin{equation}
  \mathcal{E}(t) = \left(\ln\mathcal{T}-\avg{\ln x(t)} \right)^{2}
\end{equation}
The group diversity in period $t$, $\mathcal{D}(t)$, is the variance of
the opinion distribution:
\begin{equation}
  \mathcal{D}(t) = \dfrac{1}{N}\sum_{i=1}^{N}\left(\ln x_{i}(t) -
    \avg{\ln x(t)}\right)^{2}
\end{equation}
Finally, the wisdom of crowds indicator, $\mathcal{W}(t)$, measures how
much deviation from the most central estimate is needed to encompass, or
bracket, the true value. More precisely, $\mathcal{W}(t) = \max\{i |
\hat{x}_{i}(t) \le \mathcal{T} \le \hat{x}_{N-i+1}(t)\}$, where
$\hat{x}_{i}$'s are the original (i.e. not log-transformed)
\textit{sorted} opinions. The indicator has a maximum of $N/2$ when the
truth lies between the most central estimates (or is the most central
estimate) and a minimum of zero when the truth is outside the range of
all estimates.

The experiment demonstrated that even trace amount of social influence,
in terms of knowledge about others' opinions, has a negative effect on
the wisdom of crowds. This is manifested via three effects: ``Social
influence'', ``Range reduction'' and ``Self-confidence'' effect. The first leads
to convergence of opinions, i.e., reduction of group diversity, without
improving the collective error significantly. Range reduction reveals
that the core range of estimates needed to enclose the true value
gradually increases (i.e. the wisdom of crowds indicator decreases,
meaning that the true value becomes less central in the distribution of
opinions), while at the same time the distribution becomes narrower due
to the social influence effect. As a result, the crowd slowly converges
to a wrong value. Finally, the self-confidence effect demonstrates that
individuals become increasingly confident in their opinions, whereas
concurrently the group converges away from the truth. The authors
concluded that social influence undermines the wisdom of crowds, as the
population collectively drifts away from the truth with increasing
confidence.

The agent-based model introduced below reproduces the social influence
and range reduction effects in an artificial population under the no- and
aggregate-information scenarios and, based on this framework, we will
show that the decline of crowd wisdom cannot be imputed to social
influence alone.

\section{The Model}
\label{model-description}
Consider a population of $N$ individuals, each possessing a continuous
opinion $x_{i}(t)$ at time $t$. We posit that the opinion of agent $i$
evolves according to the following process:
\begin{equation}
  \label{estimates}
  \d{}{t}x_{i}(t) = \alpha_{i} \left( \avg{x(t)} - x_{i}(t) \right) + \beta_{i}
  \left( x_{i}(0) - x_{i}(t) \right) + D \xi_{i}(t)
\end{equation}
The first term represents coupling to an individual's environment. We
refer to it as ``social influence''. In the aggregate regime, people are
only aware of the arithmetic average of all opinions, so they try to converge to that value with a given sensitivity
$\alpha_{i}$: it corresponds to the perceived strength of
social influence affecting the $i$--th individual.

The second term models individual's tendency to uphold their original
opinions. We refer to it as ``individual conviction'', and its strength
is given by the parameter $\beta_{i}$.

The third term corresponds to the fact that individuals may change their
opinion because they incorporate other information (known {\em ex ante}),
they previously disregarded. This term does not come from interactions,
but originates purely from internal mechanisms. Then, 
$\xi_{i}(t)$ is Gaussian white noise with unit variance. $D$ is the
corresponding noise intensity.

From a physicist's point of view, this dynamics resembles Brownian
particles interacting in a mean-field
scenario \cite{SchweitzerFrank2007}. In social psychology, we can also
see Eq.~(\ref{estimates}) as a formalisation of Lewin's heuristics that
the individual action is a function of idiosyncratic perception of
available information and the influence of a ``field'', that is the
influence of their environment \cite{Sansone2004}.

With these basic ingredients set-up, we investigate numerically the
evolution of the collective error, $\mathcal{E}$, group diversity,
$\mathcal{D}$, and the wisdom of crowds indicator, $\mathcal{W}$, as it depends on the
$\{\alpha,\beta\}$ parameter space. In the simplest case, we take
$\alpha_{i}=\alpha$ and $\beta_{i}=\beta$.

\section{Results}
\label{results}
For all simulations, we used the Heun/Euler method with a constant time-step
$\Delta t=0.01$:
\begin{align*}
  x_{i}(t+\Delta t) &= x_{i}(t) + \Delta t \alpha(\avg{x(t)}-x_{i}(t))+\\
  &\quad + \Delta t \beta(x_{i}(0)-x_{i}(t)) + D \sqrt{\Delta t}
  ~\text{GRND}
\end{align*}
GRND is a Gaussian random number whose mean equals 0 and standard its
deviation is one. The noise intensity, $D = 10 ^{-3}$.

To test dependence on the initial configuration, two different starting
populations were sampled from log-normal distributions with means,
$\mu_{1}=-3$ , $\mu_{2}=-2.9$, and variances,
$\sigma^{2}_{1}=\sigma^{2}_{2}=0.72$\footnote{$\mu$ and $\sigma^{2}$
  refer to arithmetic average and variance of the log-transformed
  estimates respectively.}. A noise intensity of 0.001 for the stochastic term, thus,
ensures that the impact of the noise is small, yet present.

Each simulation was run for $N=100$ agents and $T=3000$ time steps. The
latter ensures that the population eventually reaches a steady-state,
where $dx_{i}(t)/dt=0$. The equilibrium is attained when the perceived
social influence driving an individual's opinion away from its initial
value equals the strength of individual conviction trying to keep the
individual at its original estimate.

\subsection{No-information Regime}
The {\em no-information regime} is recovered by setting
$\alpha=0$. Eq.~(\ref{estimates}) then becomes:
\begin{equation}
  \label{no-information-dynamics}
  \d{}{t}x_{i}(t) = \beta\left( x_{i}(0) - x_{i}(t) \right) + D \xi(t)
\end{equation}
This is a standard Orstein-Uhlenbeck process, whose solution is formally given by:
\begin{equation*}
  x_{i}(t) = x_{i}(0)e^{-\beta t}+x_{i}(0)\left(1-e^{-\beta t}\right) +
  D\int_{0}^{t}e^{\beta (s-t)}\xi_{i}(s)ds
\end{equation*}
Therefore the time average of the individual estimates,
$\overline{x_{i}(t)}$, drifts towards $x_{i}(0)$ for large $t$.

\begin{figure}[thbp]
  \centering
  \subfigure{\includegraphics[scale=0.2]{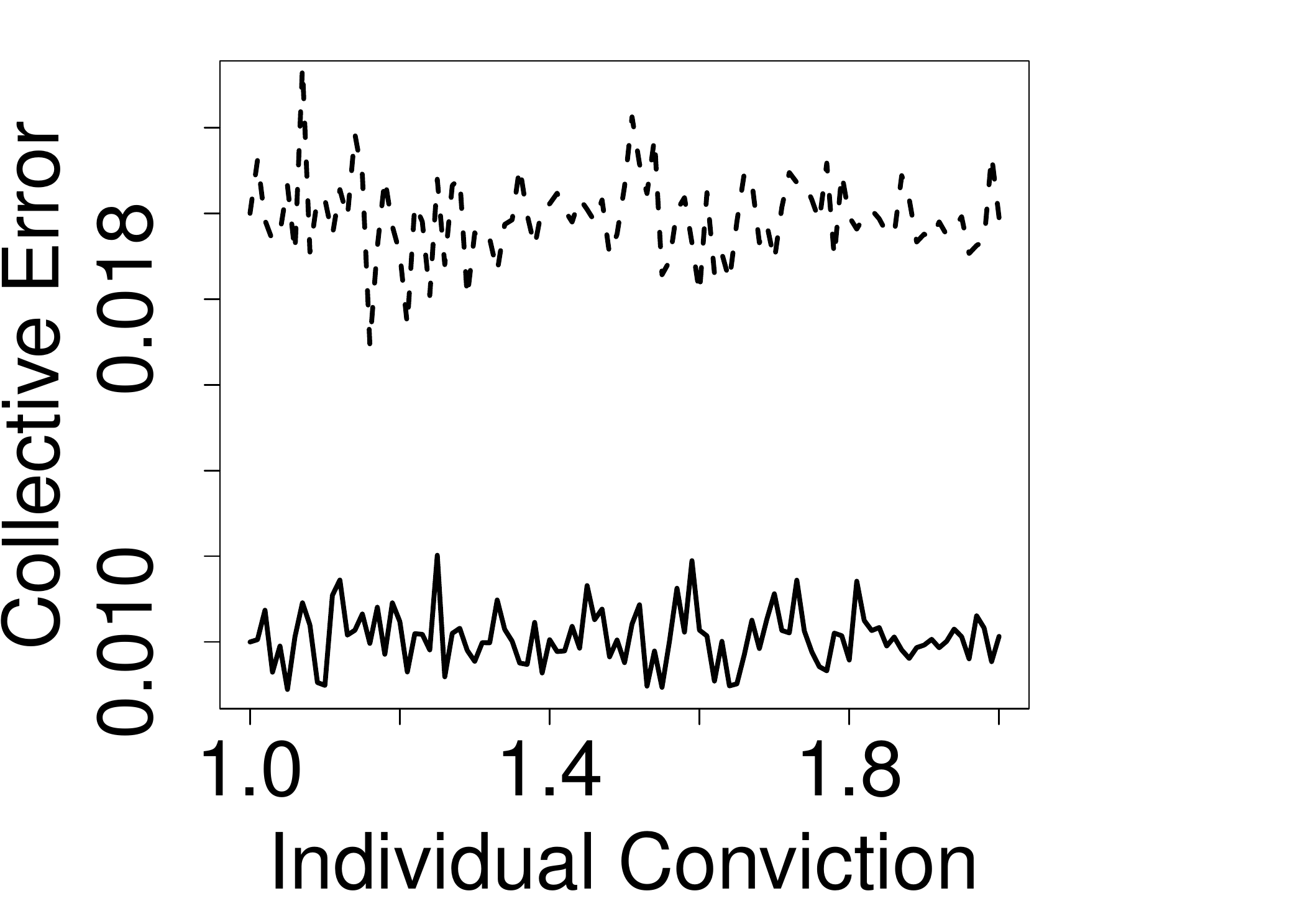}\label{ce-no-info-simu}}
  \subfigure{\includegraphics[scale=0.2]{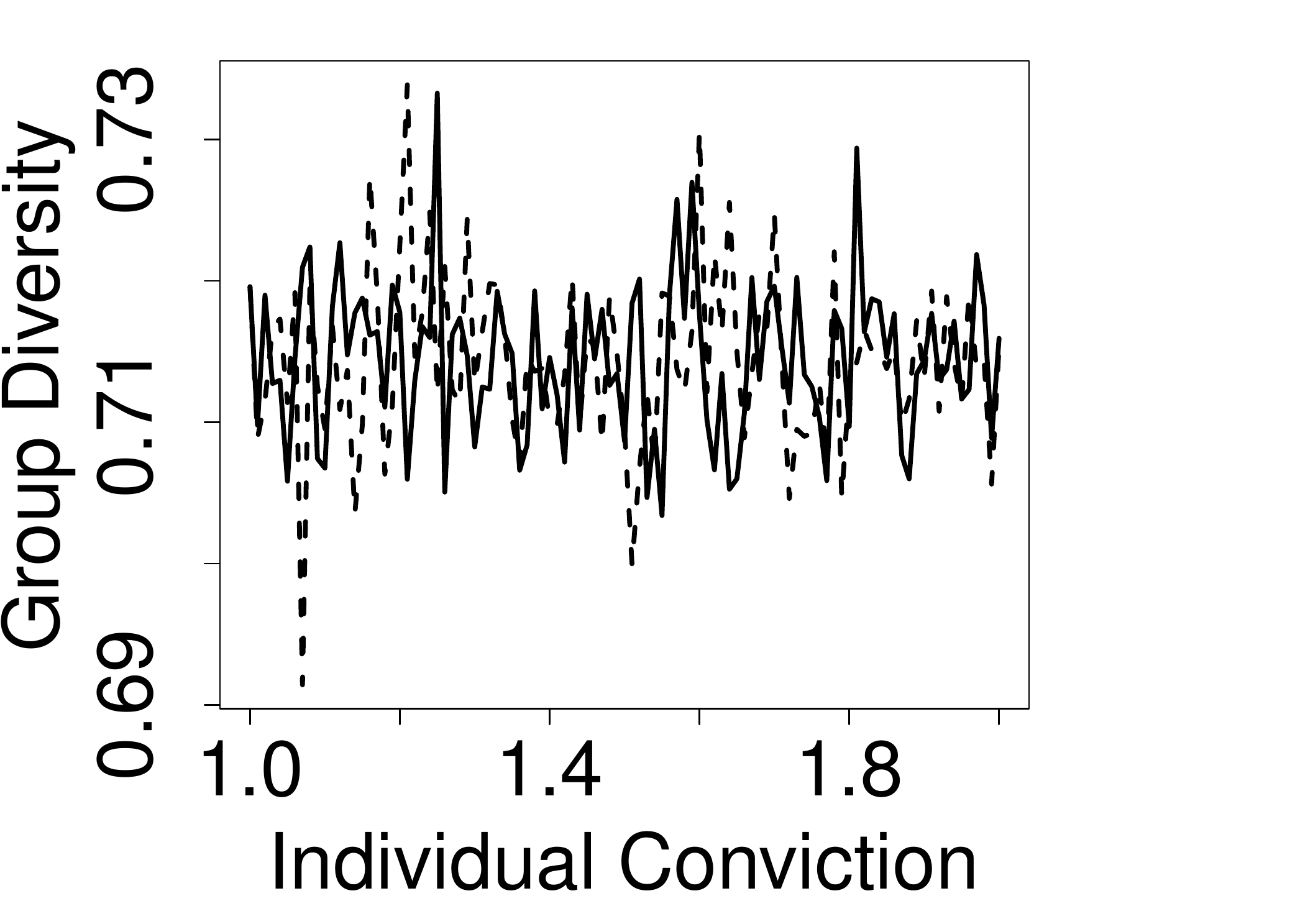}\label{diversity-no-info-simu}}\\
  \caption{Simulation of Eq.~(\ref{no-information-dynamics}). The
    aggregate measures, $\mathcal{E}$ and $\mathcal{D}$, fluctuate around
    their starting values. Left: $\mathcal{E}(0)=0.018$ (dashed),
    $\mathcal{E}(0)=0.01$ (solid). Right:
    $\mathcal{D}(0)=0.72$. Simulation parameters: $D=10 ^{-3}$.}
  \label{no-info-simu}
\end{figure}

As there is no social influence in this regime, agents do not have
incentives to deviate from their original opinions nor information on
which to base such deviations, up to small random fluctuations. In this
sense, the population can be considered static (Figure
\ref{no-info-simu}).

In Figure \ref{no-info-simu}, a simulation of two populations with
$\{\mathcal{E}_{1}(0)=0.01, \mathcal{D}_{1}(0)=0.72\}$ and
$\{\mathcal{E}_{2}(0)=0.018, \mathcal{D}_{2}(0)=0.72\}$ demonstrates that
in the long-term neither group significantly deviates from its initial
accuracy and heterogeneity. Note that the actual value of the truth,
$\ln\mathcal{T}$, is irrelevant, since of interest is only the initial
``correctness'' of the crowd, $\mathcal{E}(0)$.

Analogous result is obtained for the long-term value of $\mathcal{W}$.
\begin{figure}[htbp]
  \centering
  \includegraphics[scale=0.2]{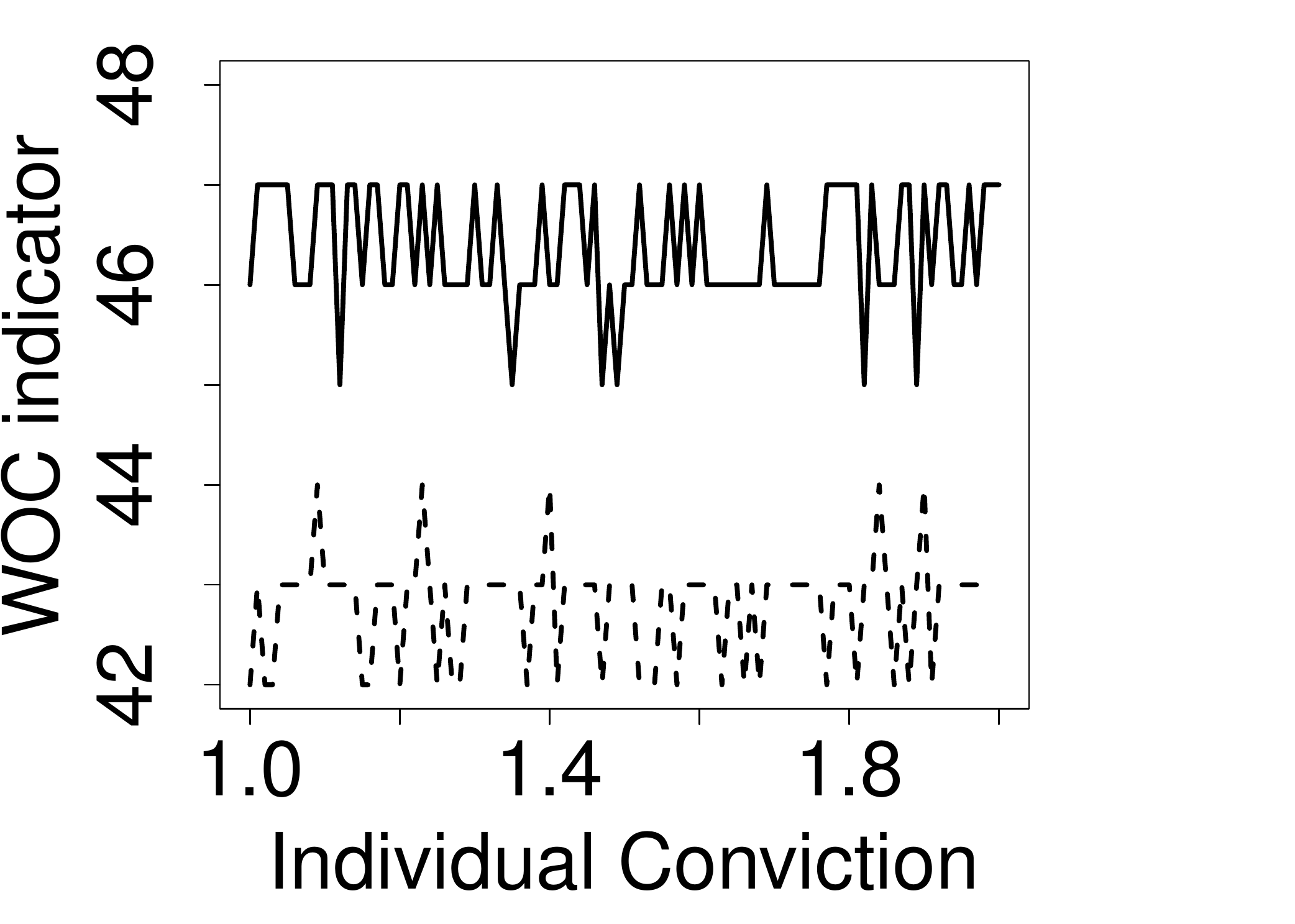}
  \caption{The Wisdom of Crowds indicator for the two populations from
    Figure \ref{no-info-simu} (same line styles). Long term behaviour is
    random fluctuations around $\mathcal{W}_{1}(0)=46$ (solid line) and
    $\mathcal{W}_{2}(0)=42$ (dashed line). Simulation parameters:
    $D=10 ^{-3}$.}
  \label{woc-no-info-simu}
\end{figure}

We conclude that in the absence of information other than one's own
judgement, groups do not tend to deviate significantly from their
original state, i.e. the wisdom of crowds is constant.

\subsection{Aggregate Information Regime}
In the aggregate information regime agents observe the average of all
estimates, which is equivalent to a mean-field scenario. Averaging
Eq.~(\ref{estimates}) over the whole population yields:
\begin{equation*}
  \d{\avg{x(t)}}{t} = \beta\left(\avg{x(0)} - \avg{x(t)}\right) + \frac{D}{\sqrt N}\avg{\xi(t)}
\end{equation*}
which is again an Ornstein-Uhlenbeck process, with solution:
\begin{align}
  \label{aggregate-information-mean}
  \avg{x(t)} &= \avg{x(0)}e^{-\beta t} + \avg{x(0)}\left(1-e^{-\beta
      t}\right) + \nonumber \\
  &\quad+\dfrac{D}{\sqrt{N}} \displaystyle \int_{0}^{t}e^{\beta(s-t)}\avg{\xi(s)}ds
\end{align}
hence, the arithmetic average is approximately constant with fluctuations
around its starting value, $\avg{x(0)}$.

Contrary to the no-information case, allowing for social interactions can
lead the group to different end states, depending on the strengths of
social influence and individual conviction. A parameter sweep in $\alpha$
and $\beta$ reveals the complete picture.

\begin{figure}[htbp]
  \centering
  \subfigure{\includegraphics[scale=0.2]{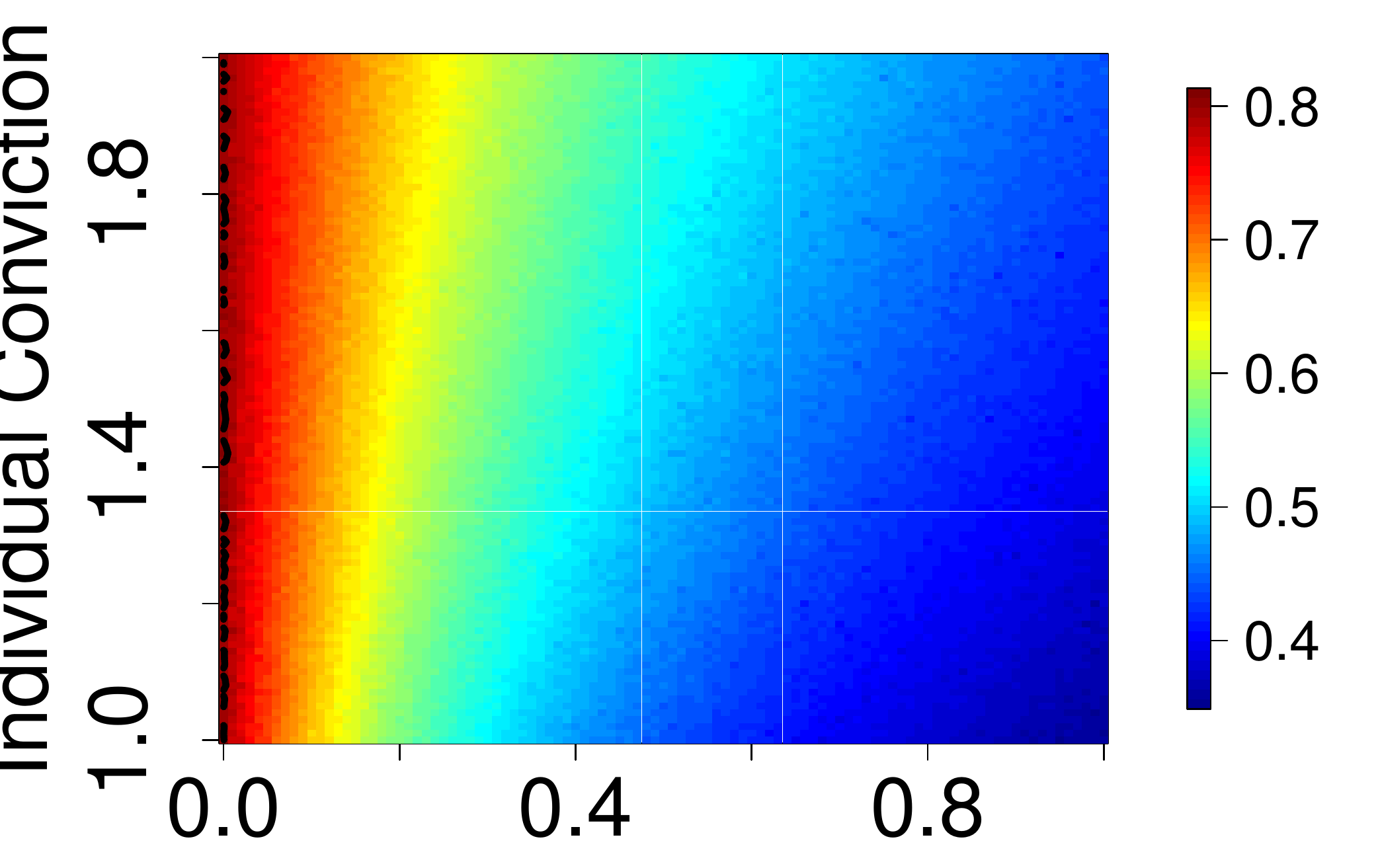}\label{cesweep4}}
  \subfigure{\includegraphics[scale=0.2]{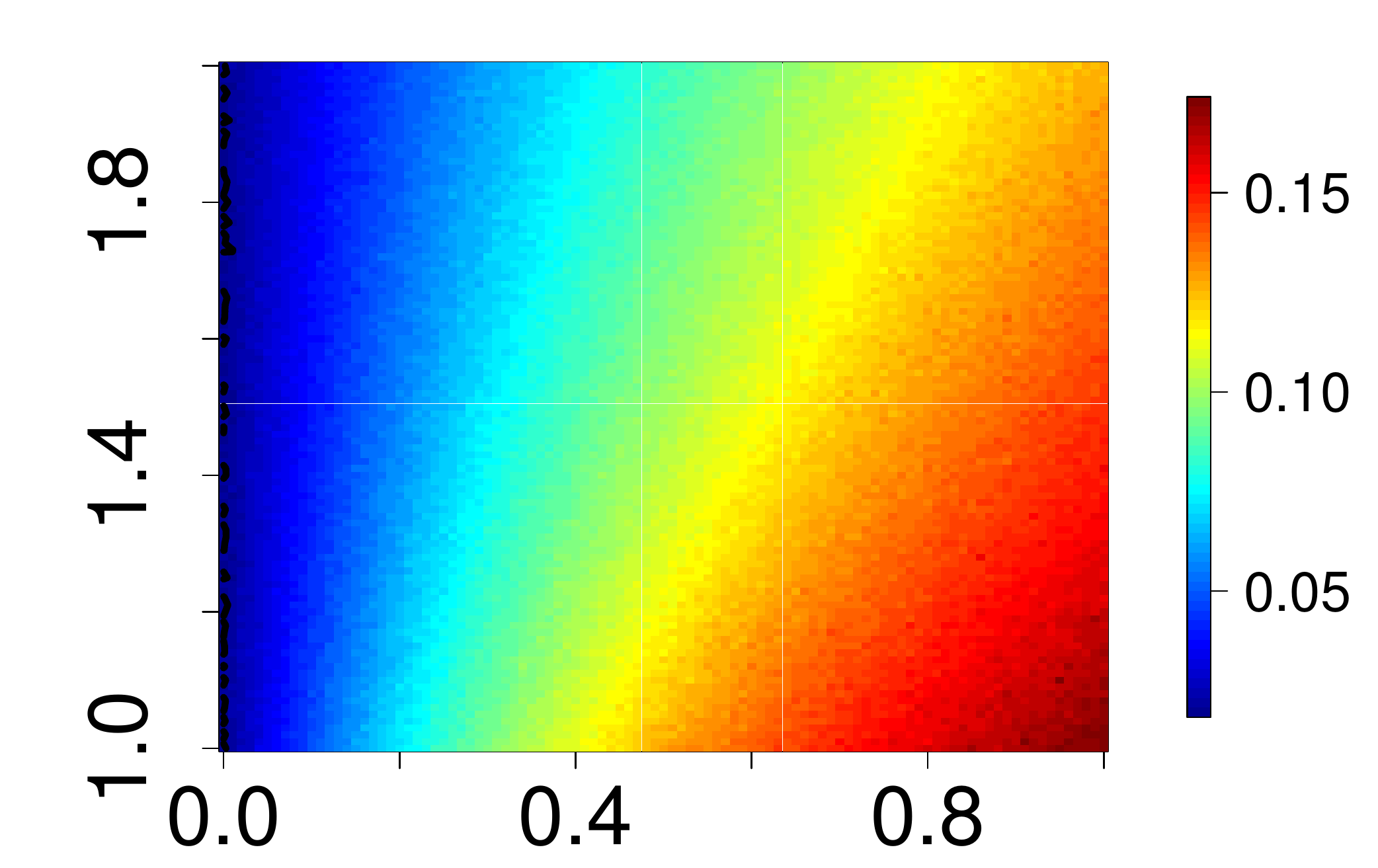}\label{cesweep5}}
  \subfigure{\includegraphics[scale=0.2]{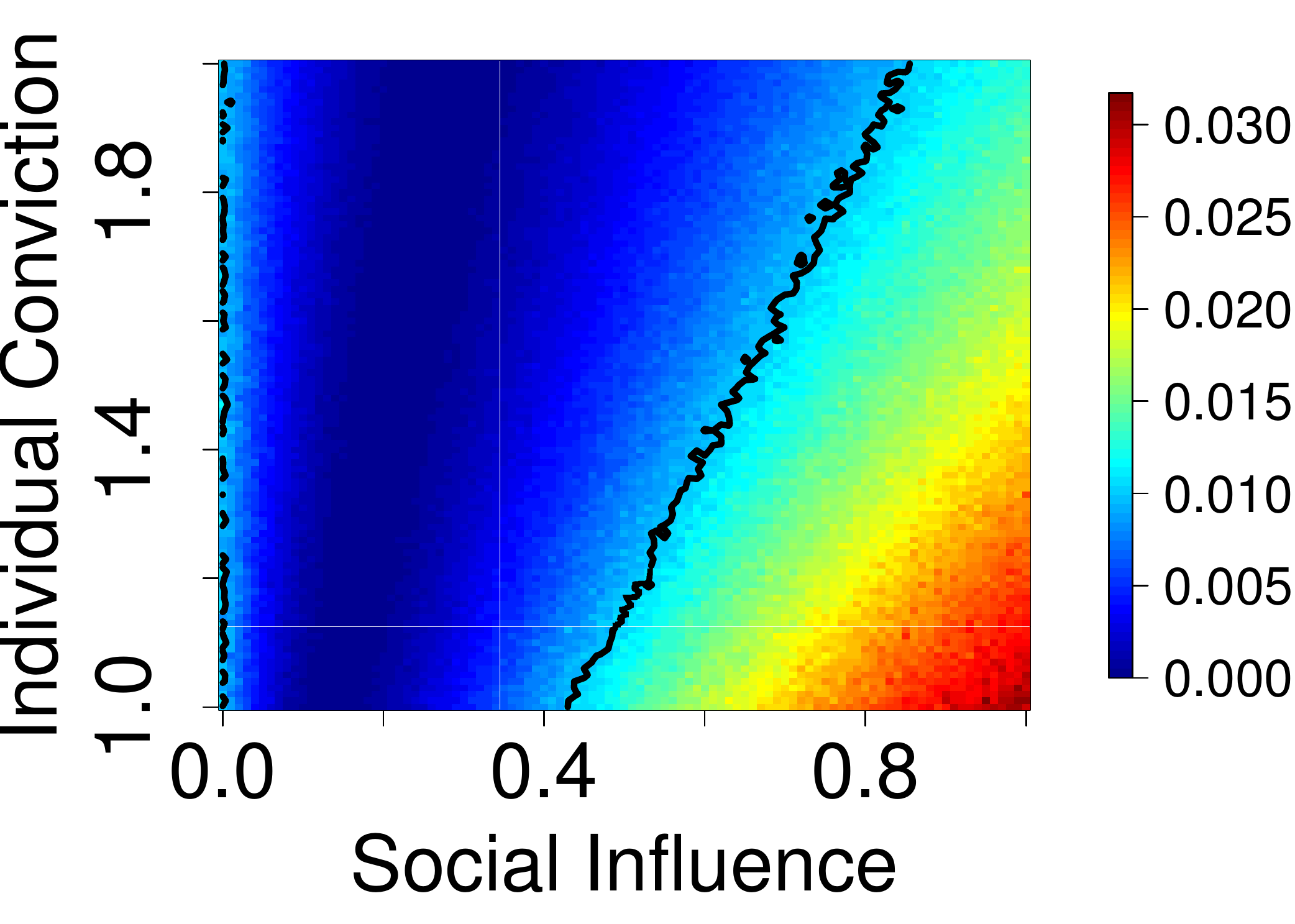}\label{cesweep6}}
  \subfigure{\includegraphics[scale=0.2]{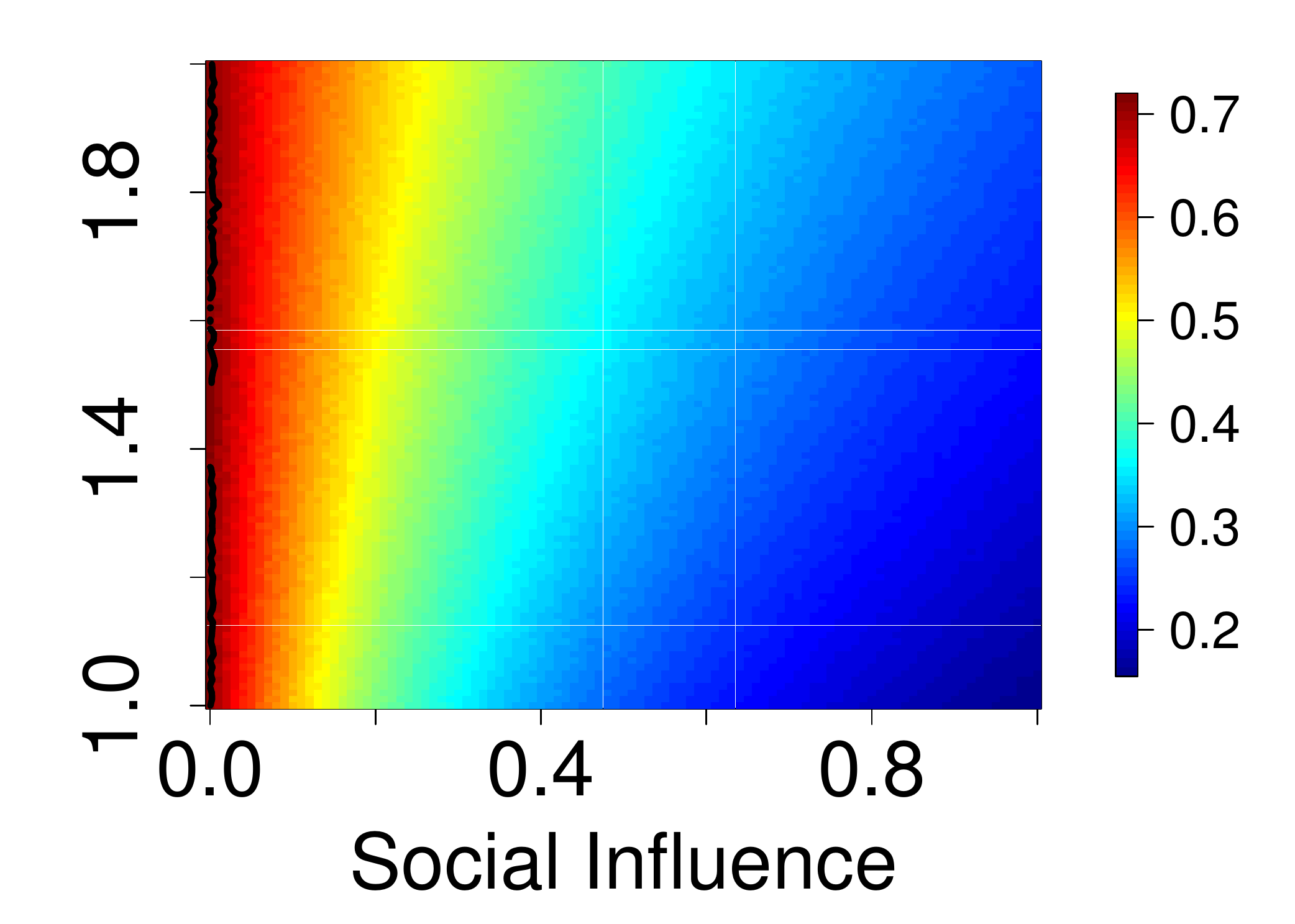}\label{diversitysweep2}}
  \caption{Ambiguous effect of social influence on the collective error
    ($\mathcal{E}$). Top-left: $\mathcal{E}(0)=0.8$, $\ln\mathcal{T}=-2$,
    $\avg{\ln x(t)}=-2.9$. Top-right: $\mathcal{E}(0)=0.02$,
    $\ln\mathcal{T}=-3.14$, $\avg{\ln x(t)}=-3$. Bottom-left:
    $\mathcal{E}(0)=0.01$, $\ln\mathcal{T}=-2.9$, $\avg{\ln
      x(t)}=-3$. The group diversity for the previous three cases is the
    same and shown in the bottom-right, with $\mathcal{D}(0)=0.72$. Black
    contour lines indicate regions where the collective error equals its
    starting value. Simulation parameters: $D=10 ^{-3}$.}
  \label{sweep}
\end{figure}

In Figure \ref{sweep} we have displayed the long-term collective errors
(top-left, top-right, bottom-left) and group diversity (bottom-right) for
three different starting configurations. The group diversity behaves
almost the
same for all three cases, hence we have shown it for the top-left
population.
Not surprisingly, agents tend to conglomerate around a common opinion in
the presence of social influence (Figure \ref{sweep}, bottom-right),
regardless of the starting configuration. We note also that individual
conviction acts in the opposite way -- it maintains diversity in the
group by reducing the perceived strength of social influence.

Figure \ref{sweep} also illustrates the opposing effects between social
influence and individual conviction -- the colour transition from blue to
red regions and vice versa. This struggle is present in all starting
configurations, however, its polarity is equivocal. In the top-left plot,
stronger individual conviction is in the group's detriment, for the
long-term collective error increases with $\beta$ for any fixed
$\alpha$. Individual conviction is beneficial in all other cases -- the
net effect is dependent on the initial state of the population.

\begin{figure}[htbp]
  \centering
  \includegraphics[scale=0.22]{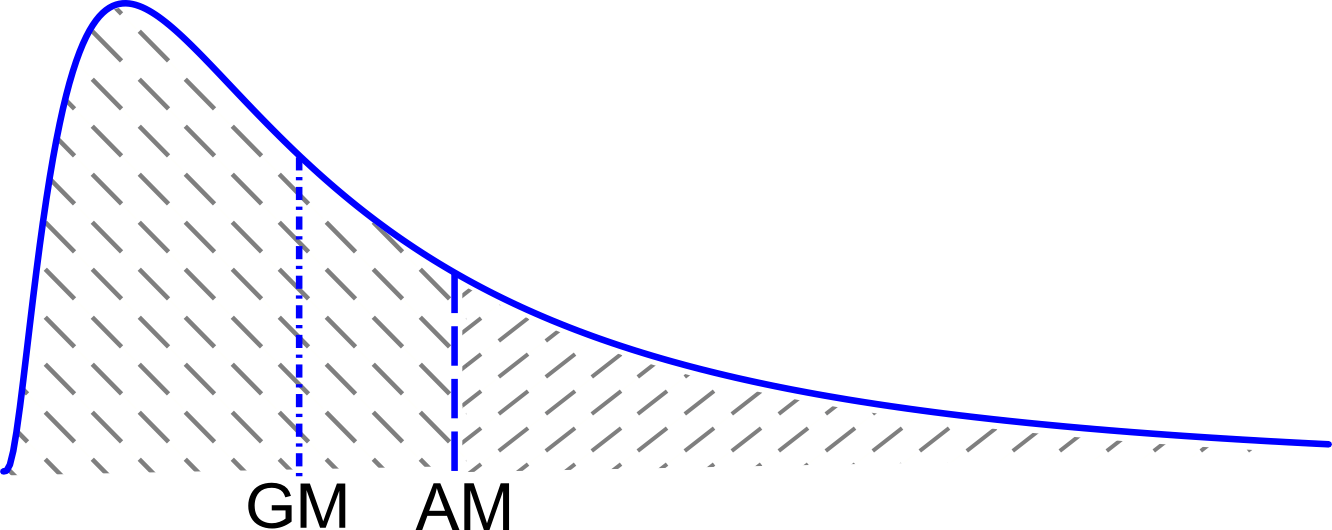}
  \caption{Rightward motion of the geometric mean (GM). The geometric
    mean equals the median (i.e. the 50th percentile) for log-normal
    distributions. However, since agents are coupled to the arithmetic
    mean (AM) and AM is greater than GM, at any given time there will be
    more mass moving to the right (negative-sloped stripes) than to the
    left (positive-sloped stripes). As a result the GM will strictly
    increase until agents reach equilibrium.}
  \label{log-normal}
\end{figure}

Further, we note that the collective error grows if $\avg{\ln
  x(0)} > \ln (\mathcal{T})$, as shown in the top-right. The reason is
that the geometric mean strictly increases when $x_{i}(t)$ is driven by
Eq.~\ref{estimates}\footnote{Thus $\ln x(t)=\ln \text{GM}$ always
  increases with time. This result can be derived analytically.}. Figure
\ref{log-normal} provides an intuition for this claim.

The top-right case in Figure \ref{sweep} essentially represents a
scenario where social influence not only does not improve the aggregate
performance of the population, but can significantly increase its
inaccuracy, reducing the group diversity at the same time.

Similar fate awaits the group even if $\avg{\ln x(0)} < \ln
(\mathcal{T})$, albeit for reduced parameter range. In the bottom-left
scenario, the aggregate opinion of the population starts from a
relatively accurate state, and ends up with a larger long-term collective
error for values of ($\alpha$, $\beta$) to the right of the second 0.01 contour
line. In effect, these two configurations reproduce the negative effect
that social influence has been shown to have in empirical studies. In
particular, \citename{Lorenz2011} describe their ``social influence effect''
as diminishing diversity in groups without improving their accuracy,
which is precisely our finding here. Therefore, despite the opposing
effect of individual conviction, for any non-zero strength of social
influence, the population ends up at a worse long-term state than the one
it started from.

Interestingly, other configurations exist where social influence brings a
clear advantage. Consider the relatively inaccurate initial population in
the top-left. Virtually for the whole parameter range, the end collective
error is lower than that in the beginning. Even more, with the weakest
individual conviction and strongest social influence, the agents actually
converge to the most accurate long-term state.

Such a positive outcome is also achieved in the bottom-left plot. Up to a
certain limit, increasing the social influence leads the population to
better stationary states\footnote{Better in the sense of lower collective
  error.} (dark blue regions), while decreasing the diversity at the same
time. Considering these cases now, one can rightly conclude the opposite
-- social influence is beneficial to the average accuracy of crowds.

Turning our attention now to the wisdom of crowds indicator,
$\mathcal{W}$, we plot its long-term behaviour for the two of the three
starting configurations in Figure \ref{sweep}.

\begin{figure}[htbp]
  \centering
  \subfigure{\includegraphics[scale=0.2]{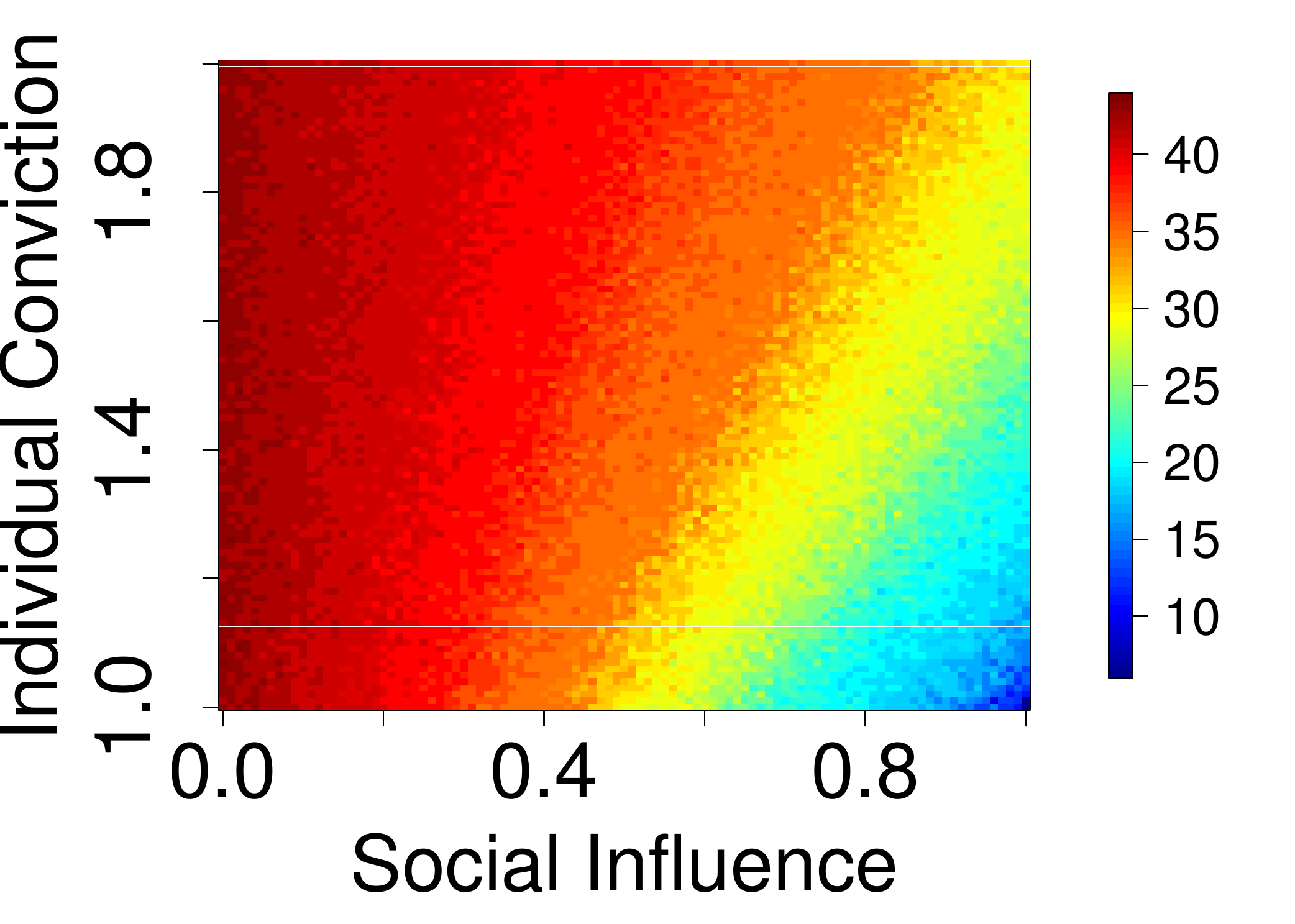}}
  \subfigure{\includegraphics[scale=0.2]{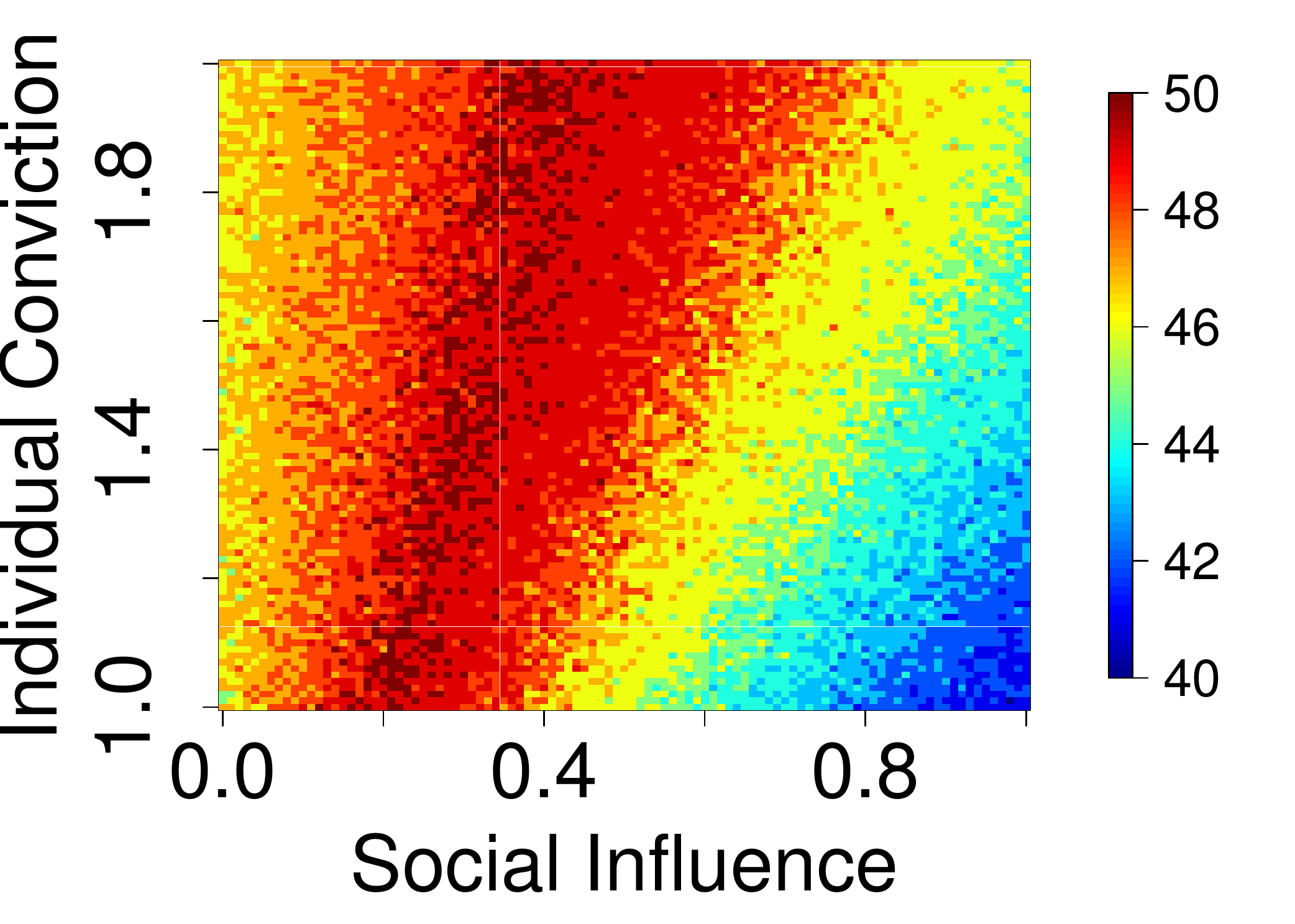}}
  \caption{Simulated long-term value of the Wisdom of Crowds
    indicator. Left: $\mathcal{W}(0)=43$, $\mathcal{E}(0)=0.02$. Right:
    $\mathcal{W}(0)=46$, $ \mathcal{E}=0.01$. Both: $\avg{\ln x(t)}=-3$,
    $\mathcal{D}(0)=0.72$. Simulation parameters as in Figure
    \ref{sweep}}
  \label{woc123-simu}
\end{figure}

The population in the left plot is the same as in Figure \ref{sweep},
top-right. Naturally, as the long-term collective error always increases
with social influence, the distribution of estimates moves farther from
the true value and gets more homogeneous at the same time. As a result,
the range of estimates needed to bracket the truth strictly increases,
which explains why $\mathcal{W}$ ends up always lower than in the
beginning. Moreover, the strength of social influence accelerates the
said drift of this group, and in turn decreases further the long-term
$\mathcal{W}$ (blue region). The same behaviour can be observed in the right
plot of Figure \ref{woc123-simu} (right-half of plot only), which is the
configuration from Figure \ref{sweep}, bottom-left -- social influence
leads the group away from the truth and decreases its wisdom, as measured
by $\mathcal{W}$. In effect, what we have described above is the ``range
reduction effect'' found in \cite{Lorenz2011} -- the truth is displaced
to peripheral regions of the opinion distribution and the group becomes
narrowly centred around a wrong value. As a consequence, a social planner
who takes a unique advise, represented by the aggregate opinion of the
group, is likely to be misled.

As with the collective error, we find that social influence can also be
beneficial for the wisdom of crowds indicator. In Figure
\ref{woc123-simu}, right, $\mathcal{W}$, actually grows with the strength
of the social interactions, and achieves the maximum value of $N/2$
(darkest red region). So, despite the loss of heterogeneity in estimates,
due to the favourable initial conditions the long-term aggregate opinion
of the group ends up closer to the true value (i.e. the collective error
decreases) for moderate amounts of social influence. As a result the
distribution of opinions is narrowly centred around the truth -- the
crowd is ``wiser''.


Explaining these ambiguous effects of social influence on $\mathcal{E}$
and $\mathcal{W}$ is one of our main results. It is not the case that
social influence is inherently ``good'' or ``bad'', and that one should
attribute group performance directly to it. Rather its impact is
modulated by the starting state of the group, in terms of
$\mathcal{E}(0)$ and $\mathcal{D}(0)$. Awareness of the latter could
allow us to determine ex-ante whether stronger social interactions are
favourable or not. In addition, the counter effect of individual
conviction could be used to remedy those populations that started from an
unfavourable state; by implementing measures, which promote individuals'
self-confidence at the expense of social influence.

\section{Conclusion}

This paper set out to study the relationship between the collective
wisdom of a group, and the social influence among the individuals
constituting it.  In particular, we aimed at explaining the ambiguous
role of social influence, which in some circumstances bears a positive
effect on the wisdom of crowds, and in others a negative one. This is a
pressing question as social interactions are practically ubiquitous --
crowds are embedded in social contexts, which invariably couple the
individuals within them. Democracies assume and rely on extensive public
discussions to form opinions and create policies. Our behaviour as
consumers, investors, voters, etc., is influenced by discussions with,
friends, colleagues and experts (among others). From a policy-maker's
perspective, this question translates to whether a government can
reliably harness the wisdom of crowds subject to heavy social
influence 
\cite{Coleman2011}. Unravelling the mechanisms (if any) by which social
influence positively or negatively affects the wisdom of crowds, becomes
then important for evaluating the trustworthiness of crowd predictions.

In a recent experiment \cite{Lorenz2011}, participants were given a
guessing task, where an objective true value had to be approached.  Each
individual updated their estimate for several rounds based on (i) no
information about others' judgements, (ii) the average estimation of the
population or (iii) full information about everyone else's estimates. The
study found that under the last two scenarios the wisdom of crowds is
weakened, which speaks for the negative effects of social influence.

In this paper, we introduced a simple model that reproduces the results
of the no- and aggregate-information scenarios in said study.  The
model consists of a population of agents endowed with a minimum set of
cognitive abilities. The agents continuously revise their estimations,
based on individual conviction (their belief in own estimations) and
social influence (from the rest of the population). We focused
on the long-term dynamics of three indicators measuring the performance
of the population: (i) the collective error, (ii) group diversity and
(iii) wisdom of crowds indicator once equilibrium has been reached.

We have demonstrated that groups whose initial average opinion is
relatively far from the truth in general benefit from stronger social
influence. The effect occurs because the stationary value the aggregated
opinion tends to (i.e.~the geometric mean in our case), and
the convergence speed to this value, increase with the strength of social influence,
which in turn reduces the collective error in the long-term. In other
words, promoting extended communication and exchange of views is more
likely to help those crowds that start off relatively wrong.

The effect of social influence, however, is detrimental to groups with a
relatively accurate initial configuration and thus suffer from an
increased drift in the aggregated opinion. In these cases, the initially
small collective error quickly reduces to zero, but then continues to
increase even beyond its starting value, due to the persistent motion of
the geometric mean. For such groups, small to moderate amounts
of information exchange are thus more beneficial.

Finally, other initial conditions exist, where even traces of social
influence leads to deterioration in the long-term collective error. This
is the main finding of \cite{Lorenz2011}, which is referred to as
``social influence'' effect.

The above discussion applies analogously to the wisdom of crowds
indicator, the quantification of how far the median estimate of the
population is from the truth. We have found starting configurations which
lead groups to be less {\em wise}\footnote{Wise, in the sense of the
  wisdom of crowds indicator.} in the long-term, for any amount of social
influence, i.e. the so-called ``range reduction'' effect from
\cite{Lorenz2011}; and configurations where groups end up wiser in the
presence of moderate social influence.

Based on these observations, our main result is that social influence in
the aggregate-regime does not directly influence the wisdom of
crowds. Rather it is the starting configuration of the population, in
terms of its collective error and group diversity, which determines the
long-term benefits or harms of social influence. The result gives insight
into how crowds can be driven to different states by modulating the strength
of social influence. For example, given some intuition about how
inaccurate the crowd initially is, we suggest that a policy-maker may
either promote social influence processes in a population or increase
individual conviction to counter undesired group influence in order to
steer the group prediction into more optimal long-term states.

It is important to stress that this result is applicable only when
individuals do not possess knowledge about the objective
truth\footnote{Except for idiosyncratic knowledge that forms initial
  opinions.}, nor do they learn or receive information that can lead them
towards it. In other words, there is no feedback between an agent's
opinion, at any given time, and their distance from the
truth. Consequently, social influence through coupling to the mean
affects relevant system-wide properties (the geometric mean in our case),
but not the collective error or wisdom of crowds indicator. As a topic
for further research, we hypothesise that even using different modelling
approaches to social influence (e.g. model the full-information regime in
\cite{Lorenz2011} or couple only to those deemed ``experts''), our
result qualitatively holds, as long as no feedback between the objective
truth and agents performance is present.

The current model assumes the existence of an objective truth, and
ignores learning, which is not the case in many real-world situations
(e.g. financial markets, political polls, etc.). However, individuals in
our model do no possess perfect knowledge of the truth, and the latter is
only present in the form of the distribution of initial guesses.
Consequently, by definition the collective error and the wisdom of crowds
indicator are driven solely by interactions and information dissemination
within the group. The true value is needed \textit{ex-post} to quantify the
current state of the population. Therefore, our proposition that it is
the crowd's starting configuration that ultimately determines the effect
of social influence can be generalised to these scenarios as well.

\label{conclusion}



\end{document}